\documentclass[preprint,showpacs,prl,twocolumn,10pt]{revtex4}%
\usepackage{amsfonts}
\usepackage{amsmath}
\usepackage{amssymb}
\usepackage{graphicx}%
\begin{document}
\preprint{ }
\title{Relaxation dynamics in quantum electron-glasses}
\author{Z. Ovadyahu}
\affiliation{Racah Institute of Physics, The Hebrew University, Jerusalem 91904, Israel }

\begin{abstract}
It is experimentally shown that, depending on the carrier-concentration of the
system $n$, the dynamics of electron-glasses either \textit{slows down }with
increasing temperature or it is \textit{independent} of it. This also
correlates with the dependence of a typical relaxation time (or `viscosity')
on $n$. These linked features are argued to be consistent with a model for
dissipative tunneling. The slow relaxation of the electron glass may emerge
then as a manifestation of friction in a many-body quantum system. Our
considerations may also explain why strongly-localized granular metals are
likely to show electron-glass effects while semiconductors are not.

\end{abstract}
\maketitle

Coupling to environment is known to have non-trivial effects on the dynamics
of quantum systems. This has been discussed in the context of a
two-state-system coupled to an oscillator-bath \cite{1}. Much less is
understood in real systems, most notably in systems that contain many, and
possibly strongly interacting two-state-systems \cite{2}. In this work we
study experimentally such a quantum system, specifically the electron-glass.
It is shown that the carrier concentration of the system plays a unique role
in determining the temperature dependence of its dynamics as well as its
`viscosity'. The results are compared with a model presented in Ref.~\cite{1}
that seems to account for the dependence of the glassy dynamics on both
temperature and carrier concentration. This, in turn, may indicate the
relevance of the orthogonality catastrophe \cite{3} to the slow relaxation in
electron glasses.

Samples used in this study were thin films of either crystalline or amorphous
indium-oxide (to be referred to as $In_{2}O_{3-x}$ and $In_{x}O$
respectively). The thickness for a particular batch (typically, 30-200~\AA )
was chosen such that at the measurement temperatures all the samples had sheet
resistance $R_{\square}$ within the range 10~M$\Omega$-100~M$\Omega.$ The
present study focused on the 4-6~K range of temperature which was achieved
either in a $^{3}$He fridge with the sample attached to a copper cold-finger,
or by employing non-ohmic fields with the sample immersed in liquid helium
\cite{4}. No difference was found in the results using either technique in
this range of temperatures.

The main method used in this work to characterize the dynamics of the electron
glasses is the double-conductance-excitation (DCE) technique more fully
described in \cite{4}, which also includes a comprehensive discussion of other
techniques for measuring dynamics and their associated caveats.%
\begin{figure}
[ptb]
\begin{center}
\includegraphics[
trim=0.000000in 4.099291in 0.000000in 1.352075in,
height=2.2589in,
width=3.0329in
]%
{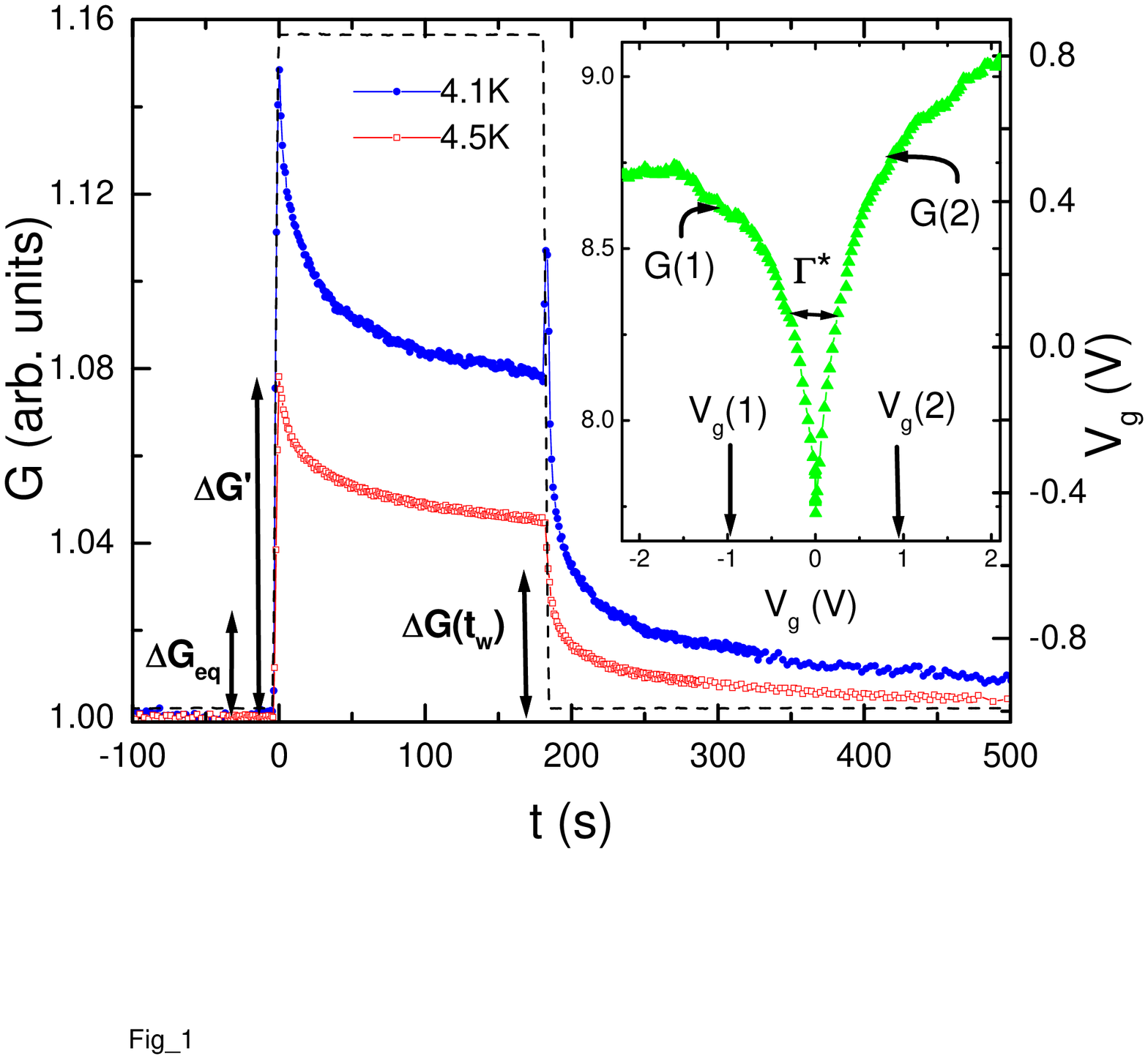}%
\caption{Typical runs of the DCE protocol at two temperatures. The dependence
of the conductance $G~$on time is plotted as squares and the gate voltage as
dashed line. Sample shown is $In_{2}O_{3-x}~$deposited on a 0.5~$\mu
m~$SiO$_{2}~$spacer thermally grown on Boron-doped Si wafer. Inset shows the
memory-dip of this sample measured in an independent set of experiments after
allowing the sample to relax for 28 hours under $V_{g}=0~$V at $T=4.1~$K. The
difference between G(2) and G(1) is the basis for estimating $\Delta G_{eq}$
(see \cite{4} for fuller details).}%
\end{center}
\end{figure}
The procedure is illustrated in Fig. 1 along with the data components that are
involved in the analysis. Starting with a voltage $V_{g}(1)$ held at the gate,
and the sample equilibrated under the fixed external conditions (temperature
$T$, or electric field $F$), one monitors the conductance as function of time
$G(t)$ to obtain the equilibrium conductance. Next, the gate voltage is
switched to $V_{g}(2)~$and is maintained there for a \textquotedblleft
waiting-time\textquotedblright\ $t_{w}~$that for all the experiments reported
here was $180\ $s. Finally, the gate voltage is switched back to $V_{g}(1),$
and $G(t)$ is measured for an additional period of time. Here we use the ratio
$\eta\equiv\Delta G(0)/\Delta G(t_{w})$ as a measure of dynamics (`viscosity')
where (\textit{c.f}., Fig. 1) $\Delta G(0)\equiv$ $\Delta G^{\prime}-\Delta
G_{eq}=G(0)-1-\Delta G_{eq};~\Delta G_{eq}\equiv\frac{G[V_{g}(2)]}%
{G[V_{g}(1)]}$ (\textit{c.f}., inset to Fig. 1)$,$ and$~\Delta G(t_{w})\equiv
G(t_{w})-1$. Note that $\Delta G(t_{w})$ depends on how far the sample
conductance has drifted in phase-space during $t_{w}~$towards its equilibrium
state (set by $V_{g}(2)$ and the external conditions, e.g., $T$). If, for
example, a full equilibrium is reached during $t_{w}$, $\Delta G(t_{w})$ will
obviously equal $\Delta G(0)~$yielding $\eta=1.$ If, on the other hand,
relaxation is infinitely slow, $\Delta G(t_{w})$ will be zero ($\eta=\infty$).
The origin of $\Delta G_{eq}$ is the thermodynamic field effect. This physical
quantity is associated with the anti-symmetric contribution to the field
effect as illustrated in the inset to Fig. 1, which also depicts the
characteristic width $\Gamma^{\ast}$ of the memory-dip \cite{5}. Note that
$\Delta G(0)$ is a proper normalization; when the external conditions are
changed, the degree of sample excitation (when $V_{g}(2)$ is switched to
$V_{g}(1)$ or vice versa) will in general change too, and it will be reflected
in $\Delta G(0).$ Likewise, the value of $\Delta G_{eq}$ is in general $\Delta
G_{eq}(T)$ and should be measured for each temperature separately. The values
of $\Delta G(0)$ and $\Delta G(t_{w})$ are extracted from the $G(t)$ data as
in Fig. 1 as follows. First, the times $t_{1}$ and $t_{2}$ where $V_{g}(2)$
reaches its final value, and $V_{g}(1)$ is reinstated respectively are noted.
These are used as the origin ($t=0)$ for the two relaxations of $G(t);$ the
first after the $V_{g}(1)$ to $V_{g}(2)$ switch, the second after the switch
back. The respective values of $G(t_{0})$ are found by extrapolation to
$t_{0~}$using the logarithmic law \cite{4} ($t_{0}$ is the resolution time of
the measurement, typically $t_{0}=1$ s), and are used to calculate $\Delta
G(0)$ and $\Delta G(t_{w})~$by subtracting the appropriate baseline
(\textit{i.e}., the equilibrium $G$~which for $\Delta G(0)~$is different than
for $\Delta G(t_{w})~$due to the anti-symmetric component shown in the inset
to Fig.~1).%
\begin{figure}
[ptbptb]
\begin{center}
\includegraphics[
trim=0.146520in 3.081551in 0.175824in 1.911944in,
height=2.5192in,
width=3.0147in
]%
{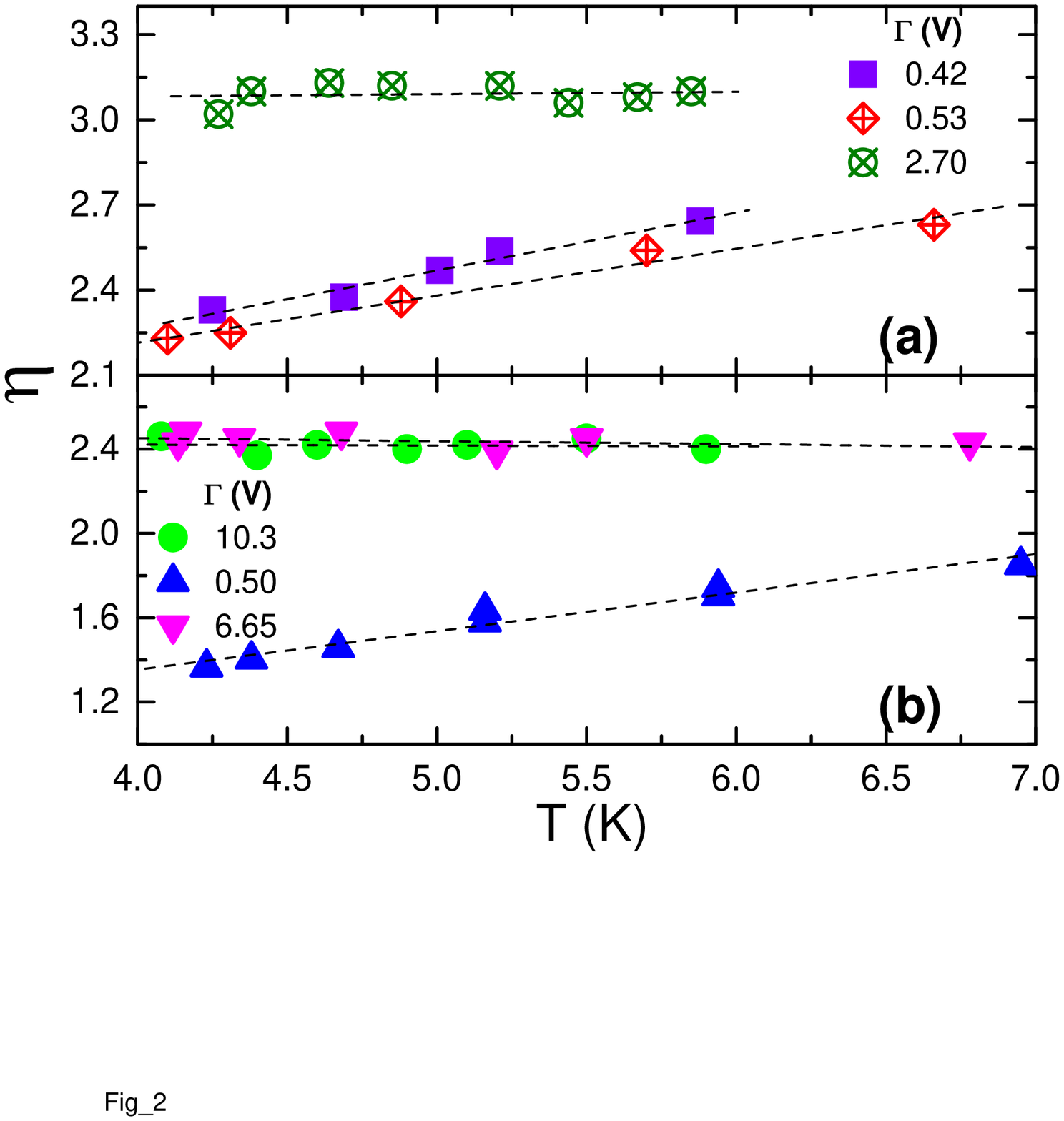}%
\caption{The dependence of the dynamics parameter $\eta~$on temperature for
typical samples with specified values of $\Gamma^{\ast}$; (a) Undoped
$In_{2}O_{3-x~}$(full squares), $In_{2}O_{3-x}~$doped with 2\% and 3.7\% Au
(diamonds and circles respectively) (b) $In_{x}O~$samples. Dashed lines are
best fits.}%
\end{center}
\end{figure}

Results for $\eta$ versus temperature based on DCE measurements are shown in
Fig. 2 for six of the studied samples. Three of these are for $In_{2}O_{3-x~}%
$films, two of which are doped with gold (Fig. 3a), and the other three are
the amorphous version (Fig. 3b). For each sample, data were taken starting at
the lowest temperature, where the sample was allowed to equilibrate for at
least 24 hours. A two hours equilibration period was used for each higher temperature.

As a further check, the dependence of the dynamics on temperature was also
assessed by measuring the excess conductance versus time $\Delta G(t)$. Such
data are available anyhow as part of the more sensitive DCE procedure
\cite{4}, and few examples for the properly normalized $\Delta G(t)$ are given
in Fig. 3.%
\begin{figure}
[ptb]
\begin{center}
\includegraphics[
trim=0.102168in 2.690549in 0.325512in 1.460875in,
height=2.6645in,
width=2.7778in
]%
{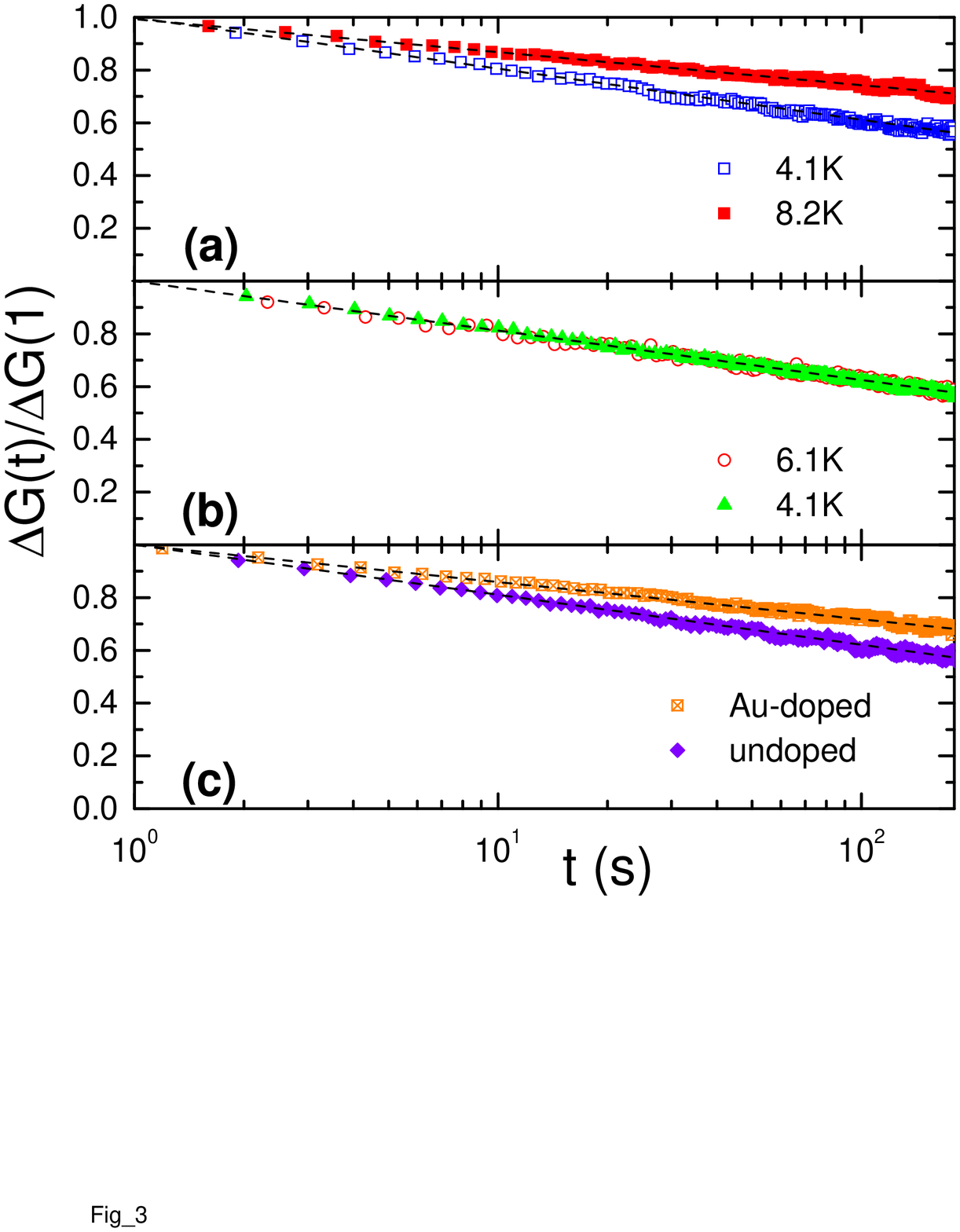}%
\caption{Normalized relaxation versus time curves for several samples. Data
are based on $\Delta G(0)~$produced by $V_{g}(1)\rightarrow V_{g}%
(2)~$excitation (\textit{c.f}., Fig.~1). (a) $In_{2}O_{3-x}~$with
$\Gamma^{\ast}=0.47~$V. (\textbf{b}) $In_{x}O~$with $\Gamma^{\ast}=6.65~$V.
(\textbf{c}) Comparing Au-doped and undoped $In_{2}O_{3-x}~$samples, both with
$R_{\square}=30~$M$\Omega,~$and both measured at 4.1~K. Dashed lines are best
linear-fits.}%
\end{center}
\end{figure}
The results of DCE measurements on the 17 samples studied in this work are
shown in Fig. 4. These suggest that, at the studied range of temperatures, the
relaxation of the electron glass either\textit{ slows down }upon\textit{
}increase of temperature\textit{, }or it is\textit{ temperature independent}.
Note that data are shown for films of $In_{2}O_{3-x},$ $In_{x}O$ (with a broad
range of $x,$ \textit{i.e}., stoichiometry), Au-doped $In_{2}O_{3-x}$, and the
granular aluminum studied by Grenet et al \cite{6}. These are quite different
systems in terms of microstructure, electron-phonon coupling, and carrier
concentrations, and they usually obey different $G(T)~$laws (though all are
activated). It is thus remarkable that their $\partial\eta/\partial T$
exhibits a unified correlation with a single parameter -- the width
$\Gamma^{\ast}$ of the memory-dip (\textit{c.f.}, inset to Fig. 1).%

\begin{figure}
[ptb]
\begin{center}
\includegraphics[
trim=0.143352in 4.678428in 0.121968in 1.304475in,
height=2.2554in,
width=3.2145in
]%
{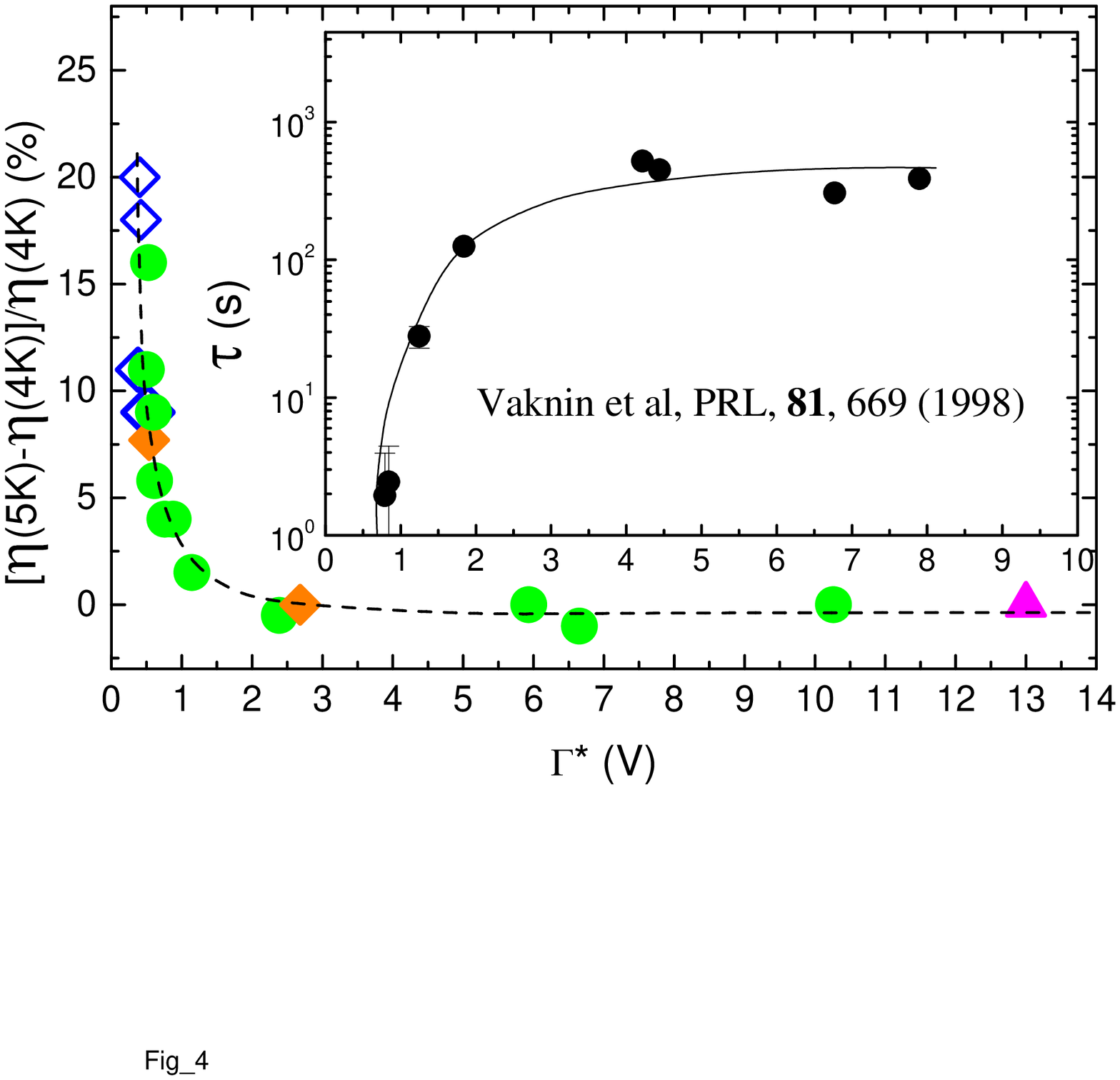}%
\caption{Temperature dependence of the relaxation dynamics vs. $\Gamma^{\ast
~}($evaluated on basis of a $0.5~\mu$m\ thick $\operatorname{Si}$O$_{2}$
spacer between the sample and the gate to allow comparison with the data in
the inset). Empty diamonds are undoped $In_{2}O_{3-x},$ full diamonds are
Au-doped $In_{2}O_{3-x},~$and circles are $In_{x}O~$samples. Dashed line is a
guide to the eye.}%
\end{center}
\end{figure}

Note that the temperature dependent dynamics is observed only for samples that
exhibit $\Gamma^{\ast}\lesssim2V.$ For small $\Gamma^{\ast}$ $\partial
\eta/\partial T$ decreases fast with $\Gamma^{\ast},$ and saturates for
$\Gamma^{\ast}\gtrsim2V$ (Fig. 4). This functional dependence of
$\frac{\partial\eta}{\partial T}[\Gamma^{\ast}]$ is strikingly similar to the
dependence of the typical relaxation time $\tau\lbrack\Gamma^{\ast}]$ measured
by the two-dip experiment in a series of $In_{x}O$ films (all with similar
values of $R_{\square})$ \cite{7}. It has been further shown by Vaknin et
al~\cite{5,7}~that $\Gamma^{\ast}$ increases monotonically with the carrier
concentration $n$ of the system that, for the range of $\Gamma^{\ast}$ in
Fig.~4, spans $n$ values from $\approx3\cdot10^{19}~cm^{-3}${\tiny \ }for
undoped $In_{2}O_{3-x}$ to $\approx10^{23}~cm^{-3}$ for granular metals. The
pattern that emerges from our results is that, all other things being equal,
the \textit{wider} is the memory-dip $\Gamma^{\ast}$ of a sample (or
equivalently, the larger is $n$), the \textit{slower} is its relaxation
(\textit{higher} `viscosity'). In addition to the systematic correlation noted
with the $In_{x}O$ data, this can be seen \textit{e.g}., in Fig. 3c comparing
the dynamics in two $In_{2}O_{3-x}$ samples with identical $R_{\square}$\ and
the sample with the larger $\Gamma^{\ast}$ exhibits slower dynamics. These
results may be then summarized as follows: Dynamics in the electron-glass is
more sluggish \textit{and} less temperature dependent the higher is the
carrier concentration $n$ ( or $\Gamma^{\ast})$ of the system. These
correlations are the basis for the main conclusions of the paper.

The linkage between temperature dependence and rate of relaxation lead us to
consider the relevance of a quantum dissipation scenario where relaxation rate
and $T$ dependence are inherently connected. To make contact with the model,
the following plausible assumptions are made: (i) The relaxation process in
the electron-glass involves hopping transitions of electrons (or group of
electrons) between localized states. Each of these may be modeled as
two-state-system, with a `bare' Rabi frequency $\Delta$. (ii) These tunneling
events interact strongly with other (localized) band electrons, and thus
coupling to the environment is Ohmic \cite{8}. (iii) The two-state-systems
that are relevant for the sluggish relaxation involve configurations with
$\varepsilon$$\ll k_{B}T$ $~$where $\varepsilon$ is the energy difference
between the two states \cite{9,10}. To have a concrete expression for the sake
of the discussion, we use a specific form of the Leggett et al model, which
gives a renormalized tunneling rate $\Delta^{\ast}$ as \cite{1}:%
\begin{equation}
\Delta^{\ast}\propto\left(  \frac{\Delta^{2}}{\omega_{c}}\right)  \frac
{\Gamma(\alpha)}{\Gamma(\alpha+\frac{1}{2})}\left[  \frac{\pi k_{B}T}%
{\hbar\omega_{c}}\right]  ^{2\alpha-1}%
\end{equation}
where $\Gamma$ is the gamma function, $\omega_{c}$ is the cut-off frequency of
the oscillators-bath that represents the environment, and $\alpha$ is the
associated coupling constant.

Eq.~1 should not be taken too literally in the context used here. The
relaxation process of the electron-glass involves a wide spectrum of tunneling
rates \cite{11}, and these events are presumably inter-dependent
(hierarchical). To characterize this convoluted dynamics by a single
`effective' $\Delta^{\ast}$ (and a single `effective' $\alpha$) as we proceed
to do next, cannot be expected to be more than a crude approximation. Other
reservations will be mentioned below. Nevertheless it seems plausible that, if
the underlying physics is relevant, the \textit{qualitative} dependence on
either $T$ or $\alpha$ predicted by Eq.~1 would be reflected in the
experimental results.

Note first and foremost that Eq. 1 allows for the uncommon $\partial
\eta/\partial T>0$ dependence observed in the limit of small $\Gamma^{\ast}%
~($\textit{i.e}$.,~$by$~\ $assigning$~\alpha\leq1/2),$ a non-trivial feature
that is hard to explain in a classical scenario. Also, the consistency of (ii)
with the data can be readily seen by interpreting the temperature dependence
of the dynamics (Fig. 4) through Eq. 1. This obviously yields $\alpha$
increasing with $\Gamma^{\ast}$. As mentioned above, $\Gamma^{\ast}$ increases
monotonically with the carrier concentration $n$ of the system \cite{7}.
Therefore it is natural to accept that the dissipative environment is the
electronic sea. Secondly, over a considerable range of $\Gamma^{\ast}~$(and
thus $n$)$,$ $\partial\eta/\partial T\approx0$. This means that $\alpha,$
while increasing with $n,$ saturates at $\approx1/2$ for large $n$.
Intriguingly, this behavior is theoretically\textit{\ }anticipated \cite{8}
for a \textit{metallic} bath where it was shown that $1/2$ is the maximum
value of $\alpha$ attainable in the high $n$ limit \cite{12}. That a similar
situation may occur in an Anderson insulator, while not impossible (at least
in the sense of $\alpha$ saturating for large $n$), remains an open question.

Another non-trivial aspect of the results is the dependence of dynamics on
$\Gamma^{\ast}$ at a given temperature. A pronounced decrease of $\Delta
^{\ast}$ with $\alpha$ is in fact what one gets from Eq. 1 for the range of
parameters relevant for these experiments. Associating $\omega_{c}$ with the
Fermi energy \cite{13} $E_{F}$ gives $\omega_{c}\approx10^{13}-5\cdot
10^{14}\sec$ for the $In_{x}O$ series. Then using Eq. 1 for $T=4~K$ results in
a sharp increase of $\Delta^{\ast}$ (by orders of magnitude) when $\alpha$
changes from 0.1 to $1/2$ (which occurs over the very narrow range of
$\Gamma^{\ast},$ \textit{c.f}., Fig~4). The correlation between the data in
Fig. 4 and the data in the inset in terms of the sharp change at the same
value of $\Gamma^{\ast}$ is then quite consistent with this picture \cite{14}.

The dramatic slowdown of the tunneling rate with $n$ naturally explains why
sluggish relaxation is peculiar to Anderson insulators with relatively high
carrier concentrations. This feature is implicit in Eq. 1 as both $\omega_{c}$
and $\alpha$ increase with the carrier concentration. The underlying mechanism
beyond these quantum `friction' effects is the Anderson orthogonality
catastrophe (AOC) \cite{3,8}, which may naturally account for the relevance of
the memory-dip $\Gamma^{\ast}${\small :} As conjectured by several authors,
the memory-dip is a reflection of an underlying density-of-states modulation
brought about by disorder and interactions \cite{15}, and the AOC is the
generic mechanism responsible for these effects \cite{16}.

The AOC is usually discussed only in the metallic regime. However, as pointed
out by Ng, the long range Coulomb interaction makes it a viable mechanism for
the strongly localized regime as well, although a real divergence may occur
only in 3D \cite{17}. At finite temperatures no divergence is expected anyhow
as the frequencies that contribute to the AOC are cut-off at $\omega
_{l}\approx k_{B}T/\hbar$. A state localized over a length $\xi\ $introduces
$\omega_{\xi}\approx\hbar/(m\xi^{2})$ as a cut-off ($m$ is a the electron
mass). Therefore, at $T\approx$ 4~K, regions of the system where the
localization length $\xi$ is larger than $\approx$10$^{2}~$\AA ~are not more
restrictive to the AOC than a metal would be at this temperature. Note that
these $\xi^{\prime}s~$contain $\gtrsim30$ electrons even for samples with the
smallest $n$ used in our studies. This presumably is the reason for the
similar electron-glass properties shared by granular metals and the
$In_{2}O_{3-x}$ and $In_{x}O$ samples and the lack of these effects in
semiconductors. Semiconductors in the localized regime typically have
$n\ll10^{20}~cm^{-3}$ we then expect that they exhibit very fast relaxation
rates, and their (correspondingly narrow) memory-dip will be anyhow masked by
the huge sensitivity of $G$ to changes in $V_{g}$ as already remarked
elsewhere \cite{18}. Sluggish relaxations observed in systems with small $n$
are probably due to coupling to a slowly varying extrinsic potential,
\textit{e.g}., structural defects.

In summary, we have shown that at liquid helium temperatures, the dynamics of
several electron-glasses exhibit some surprising similarities with the
behavior expected of a single two-state-system coupled to an electronic bath.
These include an unusual temperature dependence of dynamics, and a strong
suppression of relaxation rates caused by coupling to a dissipative bath. It
would be interesting to further test some of the conjectures raised in this
paper by carrying out similar studies at lower temperatures (where $\eta$
should become $T$ independent). On the theoretical side, the intriguing
question is how the orthogonality catastrophe and in particular the behavior
of the coupling constant $\alpha$ as function of $n\ $are affected when the
metallic system crosses over to the localized regime. These issues clearly
deserve a serious theoretical elucidation.

This research was supported by a grant administered by the US Israel
Binational Science Foundation and by the Israeli Foundation for Sciences and Humanities.


\begin{thebibliography}{99}
\bibitem {9}The excess conductance relaxation that we focus on here has
relative amplitude of typically few percents which suggests that the excess
energy in small relative to $T.$ The dc conductivity on the other hand, is
composed of hops that involve $\varepsilon\gg k_{B}T.$ In this case the
`biased' expression should be used (Eq.~7.18 in \cite{1}), which would merely
modify the values of the Miller-Abrahams resistors in the system.
\bibitem {10}This expression is valid for $T\gg\frac{\hbar\Delta}{k_{B}\alpha
}\left[  \frac{\Delta}{\omega_{c}}\right]  ^{\frac{\alpha}{1-\alpha}}$, which
is easily satisfied over the entire range of parameters used here for all
transitions with $\Delta\lesssim10^{10}~\sec.$
\bibitem {14}Note that $\tau$ defined by the two-dip-experiment is
exponentially more sensitive than either the DCE or the single relaxation
technique \cite{4}; If, (\textit{e.g.}, Fig. 3), the dynamics is assessed by
the logarithmic slope $a$ in $\Delta G=1-a\log(t),$ then the
two-dip-experiment would give $\tau\propto10^{(-1/2a)}~$(the factor 2 results
from the excitation-relaxation symmetry inherent to the two-dip-experiment,
\cite{5}).
\end{thebibliography}
\end{document}